
%
\newbox\leftpage \newdimen\fullhsize \newdimen\hstitle \newdimen\hsbody
\tolerance=1000\hfuzz=2pt
\def\printertype{ps: }
\def\qms{\def\printertype{qms: }
\ifx\answ\bigans\else\voffset=-.4truein\hoffset=.125truein\fi}
\def\bigans{b }
%
\let\answ\bigans
\ifx\answ\bigans\message{This will come out unreduced.}
\magnification=1200\baselineskip=12pt plus 2pt minus 1pt
\hsbody=\hsize \hstitle=\hsize 
\else\message{(This will be reduced.} \let\lr=L
\magnification=1000\baselineskip=16pt plus 2pt minus 1pt
\voffset=-.31truein\vsize=7truein\hoffset=-.59truein
\hstitle=8truein\hsbody=4.75truein\fullhsize=10truein\hsize=\hsbody
\output={\ifnum\pageno=0 
  \shipout\vbox{\special{\printertype landscape}\makeheadline
    \hbox to \fullhsize{\hfill\pagebody\hfill}}\advancepageno
  \else
  \almostshipout{\leftline{\vbox{\pagebody\makefootline}}}\advancepageno
  \fi}
\def\almostshipout#1{\if L\lr \count1=1 \message{[\the\count0.\the\count1]}
      \global\setbox\leftpage=#1 \global\let\lr=R
  \else \count1=2
    \shipout\vbox{\special{\printertype landscape}
      \hbox to\fullhsize{\box\leftpage\hfil#1}}  \global\let\lr=L\fi}
\fi
%
\catcode`\@=11 
\newcount\yearltd\yearltd=\year\advance\yearltd by -1900

%
%

\def\draftmode{\message{ DRAFTMODE }\def\draftdate{{\rm preliminary draft:
\number\month/\number\day/\number\yearltd\ \ \hourmin}}%
\headline={\hfil\draftdate}\writelabels\baselineskip=20pt plus 2pt minus 2pt
 {\count255=\time\divide\count255 by 60 \xdef\hourmin{\number\count255}
  \multiply\count255 by-60\advance\count255 by\time
  \xdef\hourmin{\hourmin:\ifnum\count255<10 0\fi\the\count255}}}
\def\nolabels{\def\wrlabel##1{}\def\eqlabeL##1{}\def\reflabel##1{}}
\def\writelabels{\def\wrlabel##1{\leavevmode\vadjust{\rlap{\smash%
{\line{{\escapechar=` \hfill\rlap{\sevenrm\hskip.03in\string##1}}}}}}}%
\def\eqlabeL##1{{\escapechar-1\rlap{\sevenrm\hskip.05in\string##1}}}%
\def\reflabel##1{\noexpand\llap{\noexpand\sevenrm\string\string\string##1}}}
\nolabels
%
\global\newcount\secno \global\secno=0
\global\newcount\meqno \global\meqno=1
\def\newsec#1{\global\advance\secno by1\message{(\the\secno. #1)}
\global\subsecno=0\xdef\secsym{\the\secno.}\global\meqno=1
\bigbreak\bigskip\noindent{\bf\the\secno. #1}\writetoca{{\secsym} {#1}}
\par\nobreak\medskip\nobreak}
\xdef\secsym{}
\global\newcount\subsecno \global\subsecno=0
\def\subsec#1{\global\advance\subsecno by1\message{(\secsym\the\subsecno. #1)}
\bigbreak\noindent{\it\secsym\the\subsecno. #1}\writetoca{\string\quad
{\secsym\the\subsecno.} {#1}}\par\nobreak\medskip\nobreak}
\def\appendix#1#2{\global\meqno=1\global\subsecno=0\xdef\secsym{\hbox{#1.}}
\bigbreak\bigskip\noindent{\bf Appendix #1. #2}\message{(#1. #2)}
\writetoca{Appendix {#1.} {#2}}\par\nobreak\medskip\nobreak}
%
%
\def\eqnn#1{\xdef #1{(\secsym\the\meqno)}\writedef{#1\leftbracket#1}%
\global\advance\meqno by1\wrlabel#1}
\def\eqna#1{\xdef #1##1{\hbox{$(\secsym\the\meqno##1)$}}
\writedef{#1\numbersign1\leftbracket#1{\numbersign1}}%
\global\advance\meqno by1\wrlabel{#1$\{\}$}}
\def\eqn#1#2{\xdef #1{(\secsym\the\meqno)}\writedef{#1\leftbracket#1}%
\global\advance\meqno by1$$#2\eqno#1\eqlabeL#1$$}
%
\newskip\footskip\footskip10pt plus 1pt minus 1pt 
\def\f@@t{\baselineskip\footskip\bgroup\aftergroup\@foot\let\next}
\setbox\strutbox=\hbox{\vrule height9.5pt depth4.5pt width0pt}
\global\newcount\ftno \global\ftno=0
\def\foot{\global\advance\ftno by1\footnote{$^{\the\ftno}$}}
%
\newwrite\ftfile
\def\footend{\def\foot{\global\advance\ftno by1\chardef\wfile=\ftfile
$^{\the\ftno}$\ifnum\ftno=1\immediate\openout\ftfile=foots.tmp\fi%
\immediate\write\ftfile{\noexpand\smallskip%
\noexpand\item{f\the\ftno:\ }\pctsign}\findarg}%
\def\footatend{\vfill\eject\immediate\closeout\ftfile{\parindent=20pt
\centerline{\bf Footnotes}\nobreak\bigskip\input foots.tmp }}}
\def\footatend{}
%
%
\global\newcount\refno \global\refno=1
\newwrite\rfile
\def\ref{$^{\the\refno}$\nref}
\def\nref#1{\xdef#1{$^{\the\refno}$}\xdef\rfn{\the\refno}
\writedef{#1\leftbracket#1}%
\ifnum\refno=1\immediate\openout\rfile=refs.tmp\fi%
\global\advance\refno by1\chardef\wfile=\rfile\immediate%
\write\rfile{\noexpand\item{{\rfn}.\
}\reflabel{#1\hskip.31in}\pctsign}\findarg}%
\def\findarg#1#{\begingroup\obeylines\newlinechar=`\^^M\pass@rg}%
{\obeylines\gdef\pass@rg#1{\writ@line\relax #1^^M\hbox{}^^M}%
\gdef\writ@line#1^^M{\expandafter\toks0\expandafter{\striprel@x #1}%
\edef\next{\the\toks0}\ifx\next\em@rk\let\next=\endgroup\else\ifx\next\empty%
\else\immediate\write\wfile{\the\toks0}\fi\let\next=\writ@line\fi\next\relax}}%
\def\striprel@x#1{} \def\em@rk{\hbox{}}

\def\addref#1{\immediate\write\rfile{\noexpand\item{}#1}} 
\def\startrefs#1{\immediate\openout\rfile=refs.tmp\refno=#1}
\def\xref{\expandafter\xr@f}\def\xr@f[#1]{#1}
\def\refs#1{[\r@fs #1{\hbox{}}]}
\def\r@fs#1{\edef\next{#1}\ifx\next\em@rk\def\next{}\else
\ifx\next#1\xref #1\else#1\fi\let\next=\r@fs\fi\next}
%

%
\newwrite\ffile\global\newcount\figno \global\figno=1
\def\fig{fig.~\the\figno\nfig}
\def\nfig#1{\xdef#1{fig.~\the\figno}%
\writedef{#1\leftbracket fig.\noexpand~\the\figno}%
\ifnum\figno=1\immediate\openout\ffile=figs.tmp\fi\chardef\wfile=\ffile%
\immediate\write\ffile{\noexpand\medskip\noexpand\item{Fig.\ \the\figno. }
\reflabel{#1\hskip.55in}\pctsign}\global\advance\figno by1\findarg}
\def\xfig{\expandafter\xf@g}\def\xf@g fig.\penalty\@M\ {}
\def\figs#1{figs.~\f@gs #1{\hbox{}}}
\def\f@gs#1{\edef\next{#1}\ifx\next\em@rk\def\next{}\else
\ifx\next#1\xfig #1\else#1\fi\let\next=\f@gs\fi\next}
\newwrite\lfile
{\escapechar-1\xdef\pctsign{\string\%}\xdef\leftbracket{\string\{}
\xdef\rightbracket{\string\}}\xdef\numbersign{\string\#}}

\def\writestop{\def\writestoppt{\immediate\write\lfile{\string\pageno%
\the\pageno\string\startrefs\leftbracket\the\refno\rightbracket%
\string\def\string\secsym\leftbracket\secsym\rightbracket%
\string\secno\the\secno\string\meqno\the\meqno}\immediate\closeout\lfile}}
\def\writestoppt{}\def\writedef#1{}
\def\seclab#1{\xdef #1{\the\secno}\writedef{#1\leftbracket#1}\wrlabel{#1=#1}}
\def\subseclab#1{\xdef #1{\secsym\the\subsecno}%
\writedef{#1\leftbracket#1}\wrlabel{#1=#1}}
\newwrite\tfile \def\writetoca#1{}
\def\leaderfill{\leaders\hbox to 1em{\hss.\hss}\hfill}
\def\writetoc{\immediate\openout\tfile=toc.tmp
   \def\writetoca##1{{\edef\next{\write\tfile{\noindent ##1
   \string\leaderfill {\noexpand\number\pageno} \par}}\next}}}

%
\ifx\answ\bigans
 
 \font\titlei=cmmi10 scaled\magstep3
\font\titleis=cmmi7 scaled\magstep3 \font\titleiss=cmmi5 scaled\magstep3
\font\titlesy=cmsy10 scaled\magstep3 \font\titlesys=cmsy7 scaled\magstep3
\font\titlesyss=cmsy5 scaled\magstep3 
\else
 
 \font\titlei=cmmi10 scaled\magstep4
\font\titleis=cmmi7 scaled\magstep4 \font\titleiss=cmmi5 scaled\magstep4
\font\titlesy=cmsy10 scaled\magstep4 \font\titlesys=cmsy7 scaled\magstep4
\font\titlesyss=cmsy5 scaled\magstep4 
\font\absrm=cmr10 scaled\magstep1 \font\absrms=cmr7 scaled\magstep1
\font\absrmss=cmr5 scaled\magstep1 \font\absi=cmmi10 scaled\magstep1
\font\absis=cmmi7 scaled\magstep1 \font\absiss=cmmi5 scaled\magstep1
\font\abssy=cmsy10 scaled\magstep1 \font\abssys=cmsy7 scaled\magstep1
\font\abssyss=cmsy5 scaled\magstep1 \font\absbf=cmbx10 scaled\magstep1
\skewchar\absi='177 \skewchar\absis='177 \skewchar\absiss='177
\skewchar\abssy='60 \skewchar\abssys='60 \skewchar\abssyss='60
\fi
\skewchar\titlei='177 \skewchar\titleis='177 \skewchar\titleiss='177
\skewchar\titlesy='60 \skewchar\titlesys='60 \skewchar\titlesyss='60
\ifx\answ\bigans\def\abstractfont{\footfont}\else
\def\abstractfont{\def\rm{\fam0\absrm}
\textfont0=\absrm \scriptfont0=\absrms \scriptscriptfont0=\absrmss
\textfont1=\absi \scriptfont1=\absis \scriptscriptfont1=\absiss
\textfont2=\abssy \scriptfont2=\abssys \scriptscriptfont2=\abssyss
\textfont\itfam=\bigit \def\it{\fam\itfam\bigit}
\textfont\bffam=\absbf \def\bf{\fam\bffam\absbf} \rm} \fi
\def\tenpoint{\def\rm{\fam0\tenrm}
\textfont0=\tenrm \scriptfont0=\sevenrm \scriptscriptfont0=\fiverm
\textfont1=\teni  \scriptfont1=\seveni  \scriptscriptfont1=\fivei
\textfont2=\tensy \scriptfont2=\sevensy \scriptscriptfont2=\fivesy
\textfont\itfam=\tenit \def\it{\fam\itfam\tenit}
\textfont\bffam=\tenbf \def\bf{\fam\bffam\tenbf} \rm}
%
%
\def\noblackbox{\overfullrule=0pt}
\hyphenation{anom-aly anom-alies coun-ter-term coun-ter-terms}
\def\inv{^{\raise.15ex\hbox{${\scriptscriptstyle -}$}\kern-.05em 1}}

\def\Dsl{\,\raise.15ex\hbox{/}\mkern-13.5mu D} 
\def\dsl{\raise.15ex\hbox{/}\kern-.57em\partial}

\font\bigit=cmti10 scaled \magstep1
\def\lspace{\ifx\answ\bigans{}\else\qquad\fi}
\def\lbspace{\ifx\answ\bigans{}\else\hskip-.2in\fi} 
\def\boxeqn#1{\vcenter{\vbox{\hrule\hbox{\vrule\kern3pt\vbox{\kern3pt
	\hbox{${\displaystyle #1}$}\kern3pt}\kern3pt\vrule}\hrule}}}
\def\mbox#1#2{\vcenter{\hrule \hbox{\vrule height#2in
		\kern#1in \vrule} \hrule}}  
%

\def\darr#1{\raise1.5ex\hbox{$\leftrightarrow$}\mkern-16.5mu #1}

\def\roughly#1{\raise.3ex\hbox{$#1$\kern-.75em\lower1ex\hbox{$\sim$}}}
\font\tenmss=cmss10
\font\absmss=cmss10 scaled\magstep1
\newfam\mssfam
\font\footrm=cmr8  \font\footrms=cmr5
\font\footrmss=cmr5   \font\footi=cmmi8
\font\footis=cmmi5   \font\footiss=cmmi5
\font\footsy=cmsy8   \font\footsys=cmsy5
\font\footsyss=cmsy5   \font\footbf=cmbx8
\font\footmss=cmss8
\def\footfont{\def\rm{\fam0\footrm}
\textfont0=\footrm \scriptfont0=\footrms
\scriptscriptfont0=\footrmss
\textfont1=\footi \scriptfont1=\footis
\scriptscriptfont1=\footiss
\textfont2=\footsy \scriptfont2=\footsys
\scriptscriptfont2=\footsyss
\textfont\itfam=\footi \def\it{\fam\itfam\footi}
\textfont\mssfam=\footmss \def\mss{\fam\mssfam\footmss}
\textfont\bffam=\footbf \def\bf{\fam\bffam\footbf} \rm}
\catcode`\@=12 
%
\newif\ifdraft

\noblackbox
\catcode`\@=11
\newif\iffrontpage
%
\ifx\answ\bigans
\def\titleft{\titsm}
\magnification=1200\baselineskip=12pt plus 2pt minus 1pt
%
\voffset=0.35truein\hoffset=0.250truein
\hsize=6.0truein\vsize=8.5 truein
\hsbody=\hsize\hstitle=\hsize
\else\let\lr=L
\def\titleft{\titla}
\magnification=1000\baselineskip=14pt plus 2pt minus 1pt
%
\vsize=6.5truein
\hstitle=8truein\hsbody=4.75truein
\fullhsize=10truein\hsize=\hsbody
\fi
\parskip=4pt plus 15pt minus 1pt
\font\titsm=cmr10 scaled\magstep2
\font\titla=cmr10 scaled\magstep3
\font\tenmss=cmss10
\font\absmss=cmss10 scaled\magstep1
\newfam\mssfam
\font\footrm=cmr8  \font\footrms=cmr5
\font\footrmss=cmr5   \font\footi=cmmi8
\font\footis=cmmi5   \font\footiss=cmmi5
\font\footsy=cmsy8   \font\footsys=cmsy5
\font\footsyss=cmsy5   \font\footbf=cmbx8
\font\footmss=cmss8
\def\footfont{\def\rm{\fam0\footrm}
\textfont0=\footrm \scriptfont0=\footrms
\scriptscriptfont0=\footrmss
\textfont1=\footi \scriptfont1=\footis
\scriptscriptfont1=\footiss
\textfont2=\footsy \scriptfont2=\footsys
\scriptscriptfont2=\footsyss
\textfont\itfam=\footi \def\it{\fam\itfam\footi}
\textfont\mssfam=\footmss \def\mss{\fam\mssfam\footmss}
\textfont\bffam=\footbf \def\bf{\fam\bffam\footbf} \rm}
\def\tenpoint{\def\rm{\fam0\tenrm}
\textfont0=\tenrm \scriptfont0=\sevenrm
\scriptscriptfont0=\fiverm
\textfont1=\teni  \scriptfont1=\seveni
\scriptscriptfont1=\fivei
\textfont2=\tensy \scriptfont2=\sevensy
\scriptscriptfont2=\fivesy
\textfont\itfam=\tenit \def\it{\fam\itfam\tenit}
\textfont\mssfam=\tenmss \def\mss{\fam\mssfam\tenmss}
\textfont\bffam=\tenbf \def\bf{\fam\bffam\tenbf} \rm}
\ifx\answ\bigans\def\abstractfont{\tenpoint}\else
\def\abstractfont{\def\rm{\fam0\absrm}
\textfont0=\absrm \scriptfont0=\absrms
\scriptscriptfont0=\absrmss
\textfont1=\absi \scriptfont1=\absis
\scriptscriptfont1=\absiss
\textfont2=\abssy \scriptfont2=\abssys
\scriptscriptfont2=\abssyss
\textfont\itfam=\bigit \def\it{\fam\itfam\bigit}
\textfont\mssfam=\absmss \def\mss{\fam\mssfam\absmss}
\textfont\bffam=\absbf \def\bf{\fam\bffam\absbf}\rm}\fi
%
\def\f@@t{\baselineskip10pt\lineskip0pt\lineskiplimit0pt
\bgroup\aftergroup\@foot\let\next}
\setbox\strutbox=\hbox{\vrule height 8.pt depth 3.5pt width\z@}
\def\vfootnote#1{\insert\footins\bgroup
\baselineskip10pt\footfont
\interlinepenalty=\interfootnotelinepenalty
\floatingpenalty=20000
\splittopskip=\ht\strutbox \boxmaxdepth=\dp\strutbox
\leftskip=24pt \rightskip=\z@skip
\parindent=12pt \parfillskip=0pt plus 1fil
\spaceskip=\z@skip \xspaceskip=\z@skip
\Textindent{$#1$}\footstrut\futurelet\next\fo@t}
\def\Textindent#1{\noindent\llap{#1\enspace}\ignorespaces}
\def\footnote#1{\attach{#1}\vfootnote{#1}}%

\def\foot{\attach\footsymbolgen\vfootnote{\footsymbol}}
\let\footsymbol=\star
\newcount\lastf@@t           \lastf@@t=-1
\newcount\footsymbolcount    \footsymbolcount=0
\def\footsymbolgen{\relax\footsym
\global\lastf@@t=\pageno\footsymbol}
\def\footsym{\ifnum\footsymbolcount<0
\global\footsymbolcount=0\fi
{\iffrontpage \else \advance\lastf@@t by 1 \fi
\ifnum\lastf@@t<\pageno \global\footsymbolcount=0
\else \global\advance\footsymbolcount by 1 \fi }
\ifcase\footsymbolcount \fd@f\star\or
\fd@f\dagger\or \fd@f\ast\or
\fd@f\ddagger\or \fd@f\natural\or
\fd@f\diamond\or \fd@f\bullet\or
\fd@f\nabla\else \fd@f\dagger
\global\footsymbolcount=0 \fi }
\def\fd@f#1{\xdef\footsymbol{#1}}
\def\space@ver#1{\let\@sf=\empty \ifmmode #1\else \ifhmode
\edef\@sf{\spacefactor=\the\spacefactor}
\unskip${}#1$\relax\fi\fi}
\def\attach#1{\space@ver{\strut^{\mkern 2mu #1}}\@sf}
%
\newif\ifnref
\def\rrr#1#2{\relax\ifnref\nref#1{#2}\else\ref#1{#2}\fi}
\def\ldf#1#2{\begingroup\obeylines
\gdef#1{\rrr{#1}{#2}}\endgroup\unskip}

\nreffalse
\def\refout{\footatend\immediate\closeout\rfile\writestoppt
\baselineskip=10pt\centerline{{\bf References}}\bigskip{\frenchspacing%
\parindent=16pt\escapechar=` \input refs.tmp\vfill\eject}\nonfrenchspacing}
%
\def\eqn#1{\xdef #1{(\secsym\the\meqno)}
\writedef{#1\leftbracket#1}%
\global\advance\meqno by1\eqno#1\eqlabeL#1}
\def\eqnalign#1{\xdef #1{(\secsym\the\meqno)}
\writedef{#1\leftbracket#1}%
\global\advance\meqno by1#1\eqlabeL{#1}}
%
\def\chap#1{\newsec{#1}}
\def\chapter#1{\chap{#1}}
\def\sect#1{\subsec{{ #1}}}
\def\section#1{\sect{#1}}
\def\\{\ifnum\lastpenalty=-10000\relax
\else\hfil\penalty-10000\fi\ignorespaces}
\def\note#1{\leavevmode%
\edef\@@marginsf{\spacefactor=\the\spacefactor\relax}%
\ifdraft\strut\vadjust{%
\hbox to0pt{\hskip\hsize%
\ifx\answ\bigans\hskip.1in\else\hskip .1in\fi%
\vbox to0pt{\vskip-\dp
\strutbox\sevenbf\baselineskip=8pt plus 1pt minus 1pt%
\ifx\answ\bigans\hsize=.7in\else\hsize=.35in\fi%
\tolerance=5000 \hbadness=5000%
\leftskip=0pt \rightskip=0pt \everypar={}%
\raggedright\parskip=0pt \parindent=0pt%
\vskip-\ht\strutbox\noindent\strut#1\par%
\vss}\hss}}\fi\@@marginsf\kern-.01cm}
\def\titlepage{%
\frontpagetrue\nopagenumbers\abstractfont%
\hsize=\hstitle\rightline{\vbox{\baselineskip=10pt%
{\abstractfont\pubnum}}}\pageno=0}
\frontpagefalse
\def\pubnum{}
\def\pdate{\number\month/\number\yearltd}
\def\makefootline{\iffrontpage\vskip .27truein
\line{\the\footline}
\vskip -.1truein\leftline{\vbox{\baselineskip=10pt%
{\abstractfont\pdate}}}
\else\vskip.5cm\line{\hss \tenrm $-$ \folio\ $-$ \hss}\fi}
\def\title#1{\vskip .7truecm\titlestyle{\titleft #1}}
\def\titlestyle#1{\par\begingroup \interlinepenalty=9999
\leftskip=0.02\hsize plus 0.23\hsize minus 0.02\hsize
\rightskip=\leftskip \parfillskip=0pt
\hyphenpenalty=9000 \exhyphenpenalty=9000
\tolerance=9999 \pretolerance=9000
\spaceskip=0.333em \xspaceskip=0.5em
\noindent #1\par\endgroup }
\def\autskip{\ifx\answ\bigans\vskip.5truecm\else\vskip.1cm\fi}
\def\author#1{\vskip .7in \centerline{#1}}

\def\address#1{\ifx\answ\bigans\vskip.2truecm
\else\vskip.1cm\fi{\it \centerline{#1}}}
\def\abstract#1{
\vskip .5in\vfil\centerline
{\bf Abstract}\penalty1000
{{\smallskip\ifx\answ\bigans\leftskip 2pc \rightskip 2pc
\else\leftskip 5pc \rightskip 5pc\fi
\noindent\abstractfont \baselineskip=12pt
{#1} \smallskip}}
\penalty-1000}
%

%


\def\bfone{\relax{\rm 1\kern-.35em 1}}
\def\inbar{\vrule height1.5ex width.4pt depth0pt}
\def\IC{\relax\,\hbox{$\inbar\kern-.3em{\mss C}$}}
\def\ID{\relax{\rm I\kern-.18em D}}
\def\IF{\relax{\rm I\kern-.18em F}}
\def\IH{\relax{\rm I\kern-.18em H}}
\def\II{\relax{\rm I\kern-.17em I}}
\def\IN{\relax{\rm I\kern-.18em N}}
\def\IP{\relax{\rm I\kern-.18em P}}
\def\IQ{\relax\,\hbox{$\inbar\kern-.3em{\rm Q}$}}
\def\IR{\relax{\rm I\kern-.18em R}}
\font\cmss=cmss10 \font\cmsss=cmss10 at 7pt
\def\ZZ{\relax\ifmmode\mathchoice
{\hbox{\cmss Z\kern-.4em Z}}{\hbox{\cmss Z\kern-.4em Z}}
{\lower.9pt\hbox{\cmsss Z\kern-.4em Z}}
{\lower1.2pt\hbox{\cmsss Z\kern-.4em Z}}\else{\cmss Z\kern-.4em Z}\fi}
\def\nup#1({{\it Nucl.\ Phys.}\ $\us {B#1}$\ (}
\def\plt#1({{\it Phys.\ Lett.}\ $\us  {#1}$\ (}
\def\cmp#1({{\it Comm.\ Math.\ Phys.}\ $\us  {#1}$\ (}
\def\prp#1({{\it Phys.\ Rep.}\ $\us  {#1}$\ (}
\def\prl#1({{\it Phys.\ Rev.\ Lett.}\ $\us  {#1}$\ (}
\def\prv#1({{\it Phys.\ Rev.}\ $\us  {#1}$\ (}
\def\mpl#1({{\it Mod.\ Phys.\ Let.}\ $\us  {A#1}$\ (}
\def\ijmp#1({{\it Int.\ J.\ Mod.\ Phys.}\ $\us{A#1}$\ (}
\def\tit#1|{{\it #1},\ }
%

%

\def\tilde{\widetilde}
\def\bar{\overline}
\def\us#1{\bf{#1}}

\def\Coe#1.#2.{{#1\over #2}}

\def\coe#1.#2.{\relax{\textstyle {#1 \over #2}}\displaystyle}

\def\notin{\hbox{{$\in$}\kern-.51em\hbox{/}}}

\def\cc{$^,$}

\catcode`\@=12

\ldf\HET{D.J. Gross, J.A. Harvey, E. Martinec and R. Rohm, \nup256 (1985)
253.}

\ldf\IL{L. Ib\'a\~nez and D. L\"ust, \nup382 (1992) 305.}

\ldf\JUNG{
J. Erler, D. Jungnickel and H.P. Nilles, \plt276
(1992) 303. }

\ldf\KAPLU{V. Kaplunovsky, \nup307 (1988) 145 and erratum.}

\ldf\SUSYUN{
S. Dimopoulos, S. Raby and F. Wilczek, Phys. Rev.
D24 (1981) 1681;
                 L.E. Ib\'a\~nez and G.G. Ross, \plt105 (1981) 439;
S. Dimopoulos and H. Georgi, \nup193 (1981) 375.}

\ldf\AMALDI{
G. Costa, J. Ellis, G.L. Fogli, D.V. Nanopolous and F. Zwirner, \nup297
   (1988) 244;
J. Ellis, S. Kelley and D.V. Nanopoulos, \plt249 (1990) 441;
\plt260 (1991) 131;
P. Langacker,
preprint UPR-0435T, (1990);
U. Amaldi, W. de Boer and H. F\"urstenau, \plt260 (1991) 447;
P. Langacker and M. Luo, Phys.Rev.{\bf D44} (1991) 817;
F. Anselmo, L. Cifarelli, A. Peterman and A. Zichichi,
Nouvo Cimento {\bf 104A} (1991) 1817;
R. Roberts and G.G. Ross, \nup377 (1992) 571.}

\ldf\ORBI{L. Dixon, J. Harvey, C.~Vafa and E.~Witten,
         \nup261 (1985) 651;
        \nup274 (1986) 285;
                    L.E. Ib\'a\~nez, H.P. Nilles and F. Quevedo,
\plt187 (1987) 25.}

\ldf\DKLB{L. Dixon, V. Kaplunovsky and J. Louis,
         \nup355 (1991) 649.}

\ldf\WEIG{L. Dixon, V. Kaplunovsky and J. Louis, \nup329 (1990)
            27;
D. Bailin and A. Love, \plt288 (1992) 263;
L. Ib\'a\~nez and D. L\"ust, preprint CERN-TH.6737/92, FTUAM/92/40
(1992).}

\ldf\FILQ{A. Font, L.E. Ib\'a\~nez, D. L\"ust and F. Quevedo,
           \plt245 (1990) 401.}

\ldf\FLST{S. Ferrara,
         D. L\"ust, A. Shapere and S. Theisen, \plt225 (1989) 363.}

\ldf\GAUGINO{J.P. Derendinger, L.E. Ib\'a\~nez and H.P. Nilles,
            \plt155 (1985) 65;
         M. Dine, R. Rohm, N. Seiberg and E. Witten, \plt156 (1985) 55.}

\ldf\ILR{
I. Antoniadis, J. Ellis, R. Lacaze and D.V. Nanopoulos, \plt268 (1991)
188; S. Kalara, J.L. Lopez and D.V. Nanopoulos, \plt269 (1991) 84;
           I. Antoniadis, J. Ellis, S. Kelley and
D.V. Nanopoulos, \plt271 (1991) 31;
L.E. Ib\'a\~nez, D. L\"ust and G.G. Ross,
\plt272 (1991) 251;
D. Bailin and A. Love, \plt278 (1992) 125, \plt280 (1992) 26,
\plt292 (1992) 315; M. K. Gaillard and R. Xiu, \plt296 (1992) 71.}

\ldf\GEN{L. Ib\'a\~nez, J.E. Kim, H.P. Nilles and F. Quevedo,
\plt191 (1987) 282;
J.A. Casas, E.K. Katehou and C. Mu\~noz,
\nup317 (1989) 171;
A. Font, L.E. Ib\'a\~nez, F. Quevedo and
A. Sierra, \nup331 (1990)  421.}

\ldf\CREMMER{E. Cremmer, S. Ferrara, L. Girardello and
          A. Van Proeyen, \nup212 (1983) 413.}

\ldf\DFKZ{J.P. Derendinger, S. Ferrara, C. Kounnas and F. Zwirner,
\nup372 (1992) 145.}

\ldf\LOUIS{J. Louis, preprint SLAC-PUB-5527 (1991).}

\ldf\DUAL{K. Kikkawa and M. Yamasaki, \plt149 (1984) 357;
      N. Sakai and I. Senda, Progr. Theor. Phys. {\bf 75} (1986) 692.}

\ldf\CAOV{G. Lopes Cardoso and B. Ovrut, \nup369 (1992) 351.}

\ldf\SCHEWA{A.N. Schellekens and N.P. Warner, \nup287
(1987) 317.}

\ldf\GAVA{I. Antoniadis, E. Gava and K.S. Narain, \nup383 (1992) 93.}

\ldf\CASAS{B. de Carlos, J.A. Casas and C. Mu\~noz, preprint
CERN-TH.6681/92 (1992).}

\ldf\JL{V. Kaplunovsky and J. Louis, to appear in these proceedings.}

%
%
\voffset=0.35truein\hoffset=0.250truein
\hsize=6.0truein\vsize=8.5 truein

\def\boxit#1{\vbox{\hrule\hbox{\vrule\kern3pt\vbox{\kern4pt#1\kern4pt}
\kern3pt\vrule}\hrule}}
\def\lozenge{\boxit{\hbox to 1.1pt{%
             \vrule height 1pt width 0pt \hfill}}}

\def\abstr{In this talk we discuss string consistency requirements
on four dimensional string models,
namely the cancellation of target space duality anomalies.
The analysis is
explicitly performed for (hypothetical) orbifold models
assuming the massless spectrum of the supersymmetric standard model.
In addition, some
phenomenological properties of four-dimensional strings,
like the unification of
the standard model gauge coupling constants and soft supersymmetry
breaking parameters, are investigated.
}

\font\ninerm=cmr9
\font\ninebf=cmbx9
\font\titsm=cmr10 scaled\magstep2
%
\def\pubnum{
\hbox{CERN-TH.6819/93}}
\def\pdate{
\hbox{CERN-TH.6819/93}
\hbox{February 1993}
}
\titlepage
\vskip 2.5truecm
\title{\titsm Consistency and Phenomenology of Four-Dimensional Strings}
\bigskip
\bigskip
\bigskip
\tenpoint

\font\ninerm=cmr9
\font\ninebf=cmbx9
%
\centerline{Dieter L\"ust}
\bigskip
\centerline{{\it CERN, CH 1211 Geneva 23, Switzerland}}
\bigskip
\vfil
{\centerline{\it Talk given at the }}
{\centerline{\it 26th Workshop: ``From Superstrings to Supergravity"}}
{\centerline{\it Erice - Sicily, 5-12 December 1992}}
\bigskip
\bigskip\vfil
\noindent{\tenrm \abstr}
\vskip 3.truecm
\eject
\def\pdate{}
%
\footline={\hss\tenrm\folio\hss}
\centerline{{ \ninebf CONSISTENCY AND
PHENOMENOLOGY
}}
\centerline{{ \ninebf
OF FOUR-DIMENSIONAL STRINGS
}}
\vskip.8truecm
\centerline{{\ninerm Dieter L\"ust}}
\centerline{{\it CERN, CH 1211 Geneva 23, Switzerland}}
\vskip.5truecm
\vbox{\hbox{\centerline{{\ninerm ABSTRACT}}}
{\smallskip\leftskip 3pc \rightskip 3pc \noindent \ninerm \abstr\smallskip}}

\newsec{Introduction}

Four-dimensional string theories are regarded as excellent candidates
for unification of all interactions. However it is still an enormous
challenge for string theories to make contact with low energy physics,
i.e. with the phenomena around the weak scale.
One difficulty which emerged after the discovery of the ten-dimensional
heterotic string\HET\ is the vast proliferation of consistent
models in four dimensions. Some of the four-dimensional
string models possess in fact very attractive phenomenological
features as the standard model gauge group (plus some hidden gauge
symmetry), three families of quarks and leptons (plus some
extra vector-like states), computable, semi-realistic
Yukawa couplings, etc. But unfortunately, there is so far
not a single completely realistic model; in particular there
is no model with the standard model gauge group, three families and
no extra gauge non-singlet states.

The interactions of the massless string degrees of freedom
are described by an effective field theory where
one expands the relevant terms up to a certain power in the
external momenta. The effective low-energy lagrangian
has to obey many of the symmetry properties one knows
in point particle field theory. In particular, the requirement of
the absence of gauge and gravitational
anomalies puts severe constraints on the form of the massless
fermionic spectrum of the theory.
In string theory, however, one expects that there exist
much more symmetry than in point particle field theory
due to the finite extension of the string. One known example
of enlarged ``stringy'' symmetries are the
target-space
duality symmetries\DUAL\
which describe the invariance of the string theory
under the inversion of certain length parameters.
The duality symmetries are potentially anomalous in the low-energy
field theory. These ``stringy'' duality anomalies must be cancelled
since one knows that duality symmetries are preserved in any order
of string perturbation theory. The requirement of the
absence of target-space duality anomalies provides new
constraints\IL\ on the massless string spectrum not present in any
point particle field theory.

In this talk we want to discuss whether classes of string
compactifications, where each class contains  a large number
of different four-dimensional string models, can be free of
target-space duality anomalies when assuming a certain
massless string spectrum which looks consistent from the particle
point of view. We will focus on the spectrum of the minimal
supersymmetric standard model (MSSM). This is what we call
minimal superstring compactification (MSSC).
Clearly, this investigation can be performed for any favored,
hypothetical massless string spectrum.
Furthermore we will put MSSC's under the
phenomenological test of proper coupling constant unification.
Finally we will discuss some phenomenological consequences of
MSSC's for the soft supersymmetry (SUSY) breaking parameters
in model independent way.
The material presented in this talk is largely based on ref.\IL\
where a more complete list of references can be found.


\newsec{The effective lagrangian of MSSC}

Let us define more precisely what we mean by MSSC. (Recall
that so far a MSSC was not explicitly constructed.)
It is a hypothetical four-dimensional
string model with the following properties:
(i) It has as gauge group $G=SU(3)\times SU(2)\times U(1)\times
G_{\rm hidden}$ up to the Planck mass $M_P$. (This does note
exclude the unification of this gauge group at $M_P$.)
(ii) It has $N=1$ SUSY down to the weak scale.
(iii) The massless $SU(3)\times SU(2)\times U(1)$
non-singlet spectrum is that of the MSSM. These observable
chiral $N=1$ fields are denoted by $\Phi_i=(\phi_i,\psi_i)$ with the
flavor index
$i=Q,U,D,L,E,H,\bar H$. In addition, each string compactification
has some gauge singlet states, including the dilaton chiral
superfield $S=(s,\psi_s)$ and several moduli fields $T_\alpha=
(t_\alpha,\psi_\alpha)$.
The massive states
are at $M_P$.

The interactions of the above degrees of freedom
are described by  an $N=1$ supergravity lagrangian\CREMMER.
The kinetic energies of the matter fields and of the moduli follow
from the
K\"ahler potential which can be expanded
around $\phi_i=0$:
$$K=K_0(t_\alpha,\bar t_\alpha)+K_{ij}(t_\alpha,\bar t_\alpha)
\phi_i\bar\phi_j+\dots \eqn\kp
$$
The target space duality group $\Gamma$ is given by those discrete
reparametrizations on the moduli,
$$\Gamma:\qquad t_\alpha\rightarrow \tilde t_\alpha(t_\alpha),
\eqn\repa
$$
which leave the underlying string theory, and therefore also the
effective Langrangian invariant.
Due to the moduli dependence of the low-energy couplings,
$\Gamma$ acts non-trivially on the K\"ahler potential and on the
K\"ahler metrics. First, $\Gamma$ acts on $K$ as a $U(1)$ K\"ahler
transformation like
$$K_0\rightarrow K_0+g(t_\alpha)+\bar g(\bar t_\alpha).\eqn\ktrans
$$
Second, $\Gamma$ induces a change of the matter
K\"ahler metric of the form
$$K_{ij}\rightarrow h_{il}(t_\alpha)^{-1}\bar h_{jk}(\bar t_\alpha)^{-1}
K_{lk}.\eqn\metrtra
$$
It follows that the matter fields possess a non-trivial
transformation behavior under $\Gamma$-transformations in order
to obtain  duality-invariant kinetic-energy terms for the matter
fields:
$$\phi_i\rightarrow h_{ij}(t_\alpha)\phi_j.\eqn\mattrans
$$
Due to the non-trivial action, eq.\ktrans, of $\Gamma$ on the K\"ahler
potential, the matter fermions transform with an
additional phase as
$$\lambda_a\rightarrow e^{-{1\over 4}(g-\bar g)}\lambda_a,\qquad
\psi_i\rightarrow e^{{1\over 4}(g-\bar g)}h_{ij}\psi_j.\eqn\fermtra
$$
($\lambda_a$ are the gaugino fields.)

In the following we will restrict the discussion to
symmetric
${\bf Z}_M$ and ${\bf Z}_M\times {\bf Z}_N$ orbifolds\ORBI, since
for these type of models the effective Langrangian can be constructed
in a rather explicit way.
Our formulas will be valid for a large class of
(0,2) models with non-standard
gauge embeddings and/or with the presence of Wilson lines. These
compactifications include some examples that are of phenomenological
interest since
the gauge group can be different from $E_6\times E_8$. In fact,
there exist models with standard model gauge group $G=SU(3)\times
SU(2)\times U(1)$ and three generations plus additional vector-like
matter fields\GEN.
Every orbifold of this type has three complex planes,
and each orbifold
twist $\vec\theta=(\theta_1,\theta_2,\theta_3)$ acts either
simultaneously on two or all three planes.
For simplicity, we will consider the dependence of the effective action
on the three
untwisted moduli fields $t_\alpha$  ($\alpha=1,2,3$)
which describe the sizes
of the three complex planes.
For a more general discussion see\IL.
The moduli space for the $t_\alpha$-fields
is locally given by the non-compact
coset space $\lbrack SU(1,1)/U(1)\rbrack^3$.
Since target-space duality transformations
are discrete reparametrizations, $\Gamma$ must be a discrete subgroup
of $\lbrack SL(2,{\bf R})\rbrack^3$:
$$\Gamma:\qquad t_\alpha
\rightarrow{a_\alpha t_\alpha-
ib_\alpha\over ic_\alpha t_\alpha+d_\alpha},\qquad
a_\alpha d_\alpha-b_\alpha c_\alpha=1.\eqn\modula
$$
The parameters $a_\alpha,b_\alpha,
c_\alpha,d_\alpha$ are in general a discrete set of
real numbers.
For the overall modulus $t=t_1=t_2=t_3$,
the duality group is often given by the modular
group $SL(2,{\bf Z})$ with integer parameters.
Specifically, for (2,2) or (0,2) compactifications with possibly
non-standard embedding of the twist into the gauge group $E_8\times
E_8$, but without discrete Wilson lines, the duality
group is given by $ SL(2,{\bf Z})$.
In this case, the effective lagrangian must be modular invariant
and is determined by automorphic functions of $SL(2,{\bf Z})$\FLST.
However,
for models with discrete background parameters some parts of the
modular symmetries could be broken\JUNG.
Then the parameters $a_\alpha,b_\alpha,c_\alpha,d_\alpha$ in
eq.\modula\ form a restricted set of integers, and the effective
lagrangian will contain automorphic functions of the relevant
modular subgroup.

The K\"ahler potential has a particularly simple dependence
on the moduli fields $t_\alpha$:
$$K=-\sum_{\alpha=1}^3
\log(t_\alpha+\bar t_\alpha)
+\delta_{ij}\prod_{\alpha=1}^3
(t_\alpha+\bar t_\alpha)^{n_i^\alpha}\phi_i\bar\phi_i+\cdots
\eqn\kporbi
$$
Thus each matter field is characterized by a number $n_i^\alpha$.
Using eq.\fermtra\
it follows that the normalized matter scalars and fermions
transform under $\Gamma$ transformations \modula\ as
$$
\phi_i\rightarrow\prod_{\alpha=1}^3\Biggl({-ic_\alpha\bar t_\alpha
+d_\alpha\over ic_\alpha t_\alpha+d_\alpha}\Biggr)^{
-{1\over 2}n_i^\alpha}\phi_i,\qquad
\psi_i\rightarrow\prod_{\alpha=1}^3\Biggl({-ic_\alpha
\bar t_\alpha+d_\alpha\over ic_\alpha t_\alpha+d_\alpha
}\Biggr)^{-{1\over 4}
-{1\over 2}n_i^\alpha}\psi_i.\eqn\fermorbi
$$
Therefore, the numbers $n_i^\alpha$ are called duality charges or
modular weights of the matter fields.

In principal, the modular weights $n_i^\alpha$
are undetermined parameters
in the effective lagrangian. However it is very important to realize
that the
allowed range of possible $n_i^\alpha$'s can be computed in string theory
for any standard
model field and for any class of orbifold compactification. Without
presenting any detail, let
us just state the main result of this investigation
for the overall modular weights $n_i=\sum_{\alpha=1}^3n_i^\alpha$:
$$  \eqalign{
&n\ =\ -1  \qquad {\rm (\phi\ untwisted)}  \cr
&n\ =\ -2\ -\ p\ +\ q  \qquad {\rm (\phi\ twisted\
with\ \vec\theta\ acting\ on\  all\ planes
)}\cr
&n\ =\ -1\ -\ p\ +\ q\qquad {\rm
(\phi\ twisted\  with \ \vec\theta\ acting\ on\ two\ planes )}\cr }
\eqn \unmod
$$
where $p$ ($q$) is the number of twisted oscillators with positive
(negative) chirality in the vertex operator
of $\phi_i$\IL\cc\WEIG.
Moreover, examining the modular transformation properties
of the vertex operator of the space-time supersymmetry charge, one
exactly recovers the additional K\"ahler phase for the fermions
in eqs.\fermtra,\fermorbi.

The maximal number of oscillators is limited by the requirement
that the conformal dimension $h$ of the vertex operators must be one.
Therefore $p_{\rm max}$ and $q_{\rm max}$ depend on
contribution $h_{\rm KM}$ of the Kac-Moody part, associated to the
gauge group $SU(3)\times SU(2)\times U(1)$, to the vertex operator
of each field. $h_{\rm KM}$ is  finally determined by the level
of the $SU(3)\times SU(2)\times U(1)$ Kac-Moody algebra.
The strongest constraints on $p_{\rm max}$ and $q_{\rm max}$ arise
for the lowest Kac-Moody levels
$3/5k_1=k_2=k_3=1$. Then, for the case of ${\bf Z}_3$,
the standard model fields must not have any
oscillators, and the modular weights can only be -1 and -2.
For the other orbifolds the allowed ranges of $n_i$ can be bigger;
the complete list of cases can be found in ref.\IL.

\newsec{Consistency of MSSC -- Target Space Duality Anomalies}

Consider the supersymmetric non-linear $\sigma$-model
of the moduli $T_\alpha$ coupled to  gauge and
matter fields as described in the last section.
At the one-loop level one encounters
a triangle diagram with two gauge bosons of the gauge
group $G=\prod G_a$\foot{Analogously, there is also a mixed
gravitational, $\sigma$-model anomaly with two  gravitons
and one modulus field as external legs\CAOV\cc\IL\cc\GAVA.}
and one modulus field as external legs and massless gauginos and
charged (fermionic) matter
fields circulating inside the loop.
This anomalous diagram leads,
together with the tree-level part which is given by the
dilaton/axion field $S$,
to the following (non-local)
one-loop effective supersymmetric lagrangian\DFKZ\cc\LOUIS\cc\CAOV:
$$
{\cal L}_{\rm nl}=
\sum_a\int d^2\theta{1\over 4}W^a
W^a\biggl\lbrace k_aS
-{1\over 16\pi^2}
{1\over 16}\lozenge^{-1}\bar{\cal D}\bar{\cal D}{\cal D\cal D}
\sum_{\alpha=1}^3{b'}_a^{\alpha}\log(T_\alpha+\bar T_\alpha)
\biggr\rbrace+{\rm h.c.}\eqn\nl
$$
Here $W^a$ are the Yang--Mills superfields and $k_a$ are the levels
of the $G_a$ Kac-Moody algebras.
The anomaly coefficients ${b'}_a^{i}$ look like\LOUIS\cc\DFKZ
$${b'}_a^\alpha=-C(G_a)+\sum_{\underline R_a}
T(\underline R_a)(1+2n_{\underline R_a}^\alpha ).\eqn\coef
$$
Writing the expression \nl\ in components it leads
to a non-local contribution to the $CP$ odd term $F_{\mu\nu}
\tilde F_{\mu\nu}$ and
to a local contribution
to the gauge coupling constant.

Now, it is easy to recognize that ${\cal L}_{\rm nl}$, eq.\nl,
is not invariant under the discrete target space
duality transformation \modula.
It follows that the duality anomalies
must be cancelled by adding
new terms to the effective action.
Specifically, there are two ways of cancelling these anomalies.
In the first one\DFKZ\cc\CAOV\
the $S$ field may transform
non-trivially under duality transformations,
$S\rightarrow S-{1\over 8\pi^2}\sum_{\alpha=1}^3
\delta_{\rm GS}^\alpha\log(ic_\alpha T_\alpha+d_\alpha)$,
and cancels in this
way some part or all of the duality non-invariance of eq.\nl.
This non-trivial transformation behavior
of the $S$--field is completely analogous to the
Green--Schwarz mechanism for the case of an anomalous $U(1)$ gauge group
and leads to a mixing between the moduli and the $S$--field in the
one-loop K\"ahler potential.

Second, the target space duality anomaly can be
possibly cancelled by a local contribution to ${\cal L}_{\rm nl}$,
which is related to the one-loop threshold contributions
to the
gauge coupling constants
due to the massive string states.
The threshold contributions are given in
terms of automorphic functions of the target space duality group,
which have the required transformation behavior under the
discrete duality transformations.
This topic will be discussed in section 4.

It is clear that the part of the duality anomaly which is
removed by the Green--Schwarz  mechanism is universal, i.e.
gauge group independent.
Thus for cases where there are no moduli dependent
threshold contributions from the massive states
the anomaly coefficients have to coincide for each
gauge group factor $G_a$:
$${ {{b'}_a^\alpha}\over {k_a}} ={{{b'}_b^\alpha}
\over {k_b}}= {{{b'}_c^\alpha}\over
{k_c}} =...   \eqn\bequal $$
This particularly interesting constraint arises
for orbifold compactifications where the moduli
dependent threshold contributions are absent because of an enlarged
$N=4$ supersymmetry in the massive spectrum.
In more technical terms, eq.\bequal\ applies
if $all$ orbifold twists $\vec\theta$,
which define a particular ${\bf Z}_M$ or ${\bf Z}_M
\times{\bf Z}_N$ orbifold, act non-trivially on the corresponding
$\alpha$th
complex plane of the underlying six-torus.

Let us investigate the cancellation of duality anomalies for the
phenomenologically most interesting case of MSSC. Then
the anomaly coefficients of the standard model gauge groups
take the following form:
$$\eqalign{{b'}_3^\alpha
=3+&\sum_{g=1}^3(2n_{Q_g}^\alpha+n_{U_g}^\alpha+n_{D_g}
^\alpha),\qquad
{b'}_2^\alpha=5+\sum_{g=1}^3(3n_{Q_g}^\alpha
+n_{L_g}^\alpha)+n_{H}^\alpha+n_{\bar H
}^\alpha,\cr
{b'}_1^\alpha&=11+\sum_{g=1}^3({1\over 3}n_{Q_g}^\alpha
+{8\over 3}n_{U_g}^\alpha+{2\over 3}n_{D_g}^\alpha+n_{L_g}^\alpha+2
n_{E_g}^\alpha)+n_H^\alpha+n_{\bar H}^\alpha.\cr}\eqn\bprstand
$$
Whether eqs.\bequal\ have any solutions crucially depends
on the distribution of the allowed modular weights of the standard
model fields.
Of course, similar constraints may be obtained for other extended
gauge groups and particle contents.

The strongest constraints arise for the ${\bf Z}_3$ and ${\bf Z}_7$
orbifolds where eq.\bequal\ must be satisfied  with
respect to all three complex planes, and where the choice of
possible values for the modular weights is very limited.
Let us investigate the most
common case of level one Kac--Moody algebras,
$k_1=5/3$, $k_2=k_3=1$. (For ${\bf Z}_3$ orbifolds,
our results are also true for arbitrary $k_1$.)
With the help of a computer program one can
now check that the
equations \bequal, together with eq.\bprstand, have no simultaneous
solutions at all.
In this way we have ruled out the MSSC
${\bf Z}_3$ and ${\bf Z}_7$
compactifications with lowest Kac-Moody level
by general consistency arguments.
The requirement of target space anomaly freedom forces us to
introduce additional fields.
For ${\bf Z}_3$  we have analyzed the anomaly conditions
further. It turns out that one needs 12 more $SU(2)$
doublets that $SU(3)$ triplets in all models. This excludes in
particular the minimal Higgs content.

For all other classes of orbifolds the duality anomaly matching
conditions are unfortunately not strong enough to rule
out the MSSC scenario.
The reasons are the enlarged ranges of possible modular weights.
In addition,
all other models have at least one complex plane which
is left unrotated by one of the orbifolds twists. Then the
anomaly cancellation condition eq.\bequal\ cannot be applied with
respect to this particular complex plane.

\newsec{Phenomenology of MSSC}

\subsec{Threshold Effects and Gauge Coupling Unification}

In the last section we have demonstrated that string consistency
requirements like the cancellation of target space duality
anomalies put severe constraints on the massless spectrum
of four-dimensional strings. Now we use
the phenomenological constraint of proper
unification of the running gauge coupling constants
to get further information on the massless particle
spectrum.

The one-loop  running gauge coupling constants in four-dimensional
strings
take the following form\DKLB:
$${1\over g_a^2(\mu)}={k_a\over g^2_{\rm string}}+{b_a\over
16\pi^2}\log{M_{\rm string}^2\over\mu^2}-{1\over 16\pi^2}
\sum_{\alpha=1}^3({b'}_a^\alpha
-k_a\delta_{GS}^\alpha)\log\lbrack(t_\alpha+\bar t_\alpha)
|\Delta(t_\alpha)|^2\rbrack.\eqn\running
$$
$b_a$ is the $N=1$ $\beta$-function coefficient,
and  $M_{\rm string}$ is the typical string scale, which is
of the order of the Planck mass. Its precise value, using the
${\overline{MS}}$ scheme,
is determined by
the universal string coupling constant $g_{\rm string}$ as\KAPLU\
$M_{\rm string}=0.5\times g_{\rm string}\times
10^{18}{\rm GeV}\simeq 3.5\times 10^{17}{\rm GeV}$.
The moduli dependent term in eq.\running\ describes the
one-loop threshold contribution to the gauge coupling constant
from the massive momentum and winding modes.
Here we have assumed that the field independent contributions
to the threshold corrections are small compared to the moduli
dependent pieces. This was shown to be true for the (2,2)
${\bf Z}_3$-orbifold
compactification in ref.\KAPLU.
$\Delta (t_\alpha)$ is an automorphic function of the duality group
$\Gamma$.
Duality invariance of $g_a^2(\mu)$ requires that
the function $\Delta(t_\alpha)$
must transform (up to a phase) under \modula\ as
$\Delta(t_\alpha)\rightarrow
\Delta(t_\alpha)(ic_\alpha t_\alpha+d_\alpha)$.
For $\Gamma_\alpha=SL(2,{\bf Z})$, $\Delta(t_\alpha)$ is given by the
Dedekind function, $\Delta(t_\alpha)=\eta(t_\alpha)^2$,
where $\eta(t)=e^{-\pi t/12}\prod_{n=1}^\infty
(1-e^{2\pi nt})$.
Note that the threshold contribution is vanishing if
$t_\alpha$ corresponds to a completely rotated complex
plane, since in this case one has ${b'}_a^\alpha =k_a\delta_{GS}^\alpha$.
Thus for the ${\bf Z}_3$ and ${\bf Z}_7$ orbifolds
there are no moduli dependent threshold corrections.

Now we want to investigate the question whether the ${\bf Z}_M$ or
${\bf Z}_M\times {\bf Z}_N$ orbifold compactifications can
lead to the correct unification of the three coupling constants
of the standard model gauge group $SU(3)\times SU(2)\times U(1)_Y$,
taking into account the threshold correction of the massive string
excitations\ILR\cc\IL. Our analysis will be based on the experimentally
measured values of the strong coupling constant and the weak
mixing angle: $\alpha_3^{\rm exp}=0.115\pm 0.007$,
$\sin^2\theta_W^{\rm exp}=0.233\pm 0.0008$.
Considering the
effect of the spectrum of the minimal
supersymmetric standard model on the one-loop renormalization
group equations\SUSYUN\ without any threshold corrections
with a SUSY threshold close
to the weak scale, one finds\AMALDI\
that the quoted results for $\alpha_3$
and $\sin^2\theta_W$ are in very good agreement with a unification
mass $M_X\simeq 10^{16}{\rm GeV}$.
Comparing this value with the string scale
$M_{\rm string}
\simeq 3.5\times 10^{17}{\rm GeV}$
one finds a discrepancy.
Thus the natural question is whether the correct unification of the
three gauge coupling constants can be achieved by taking into account
threshold contributions
to $g_a^2(\mu)$.
We will focus on the MSSC scheme. Making use of eq.\running\
one gets for the value of the electroweak angle $\theta_W$
and for the strong coupling constant $\alpha_3$
($k_3=k_2=3/5k_1=1$):
$$\eqalign{
{\sin^2\theta _W}(\mu) & =
{3\over 8} -
{{3\alpha_e(\mu)}\over {32\pi}}\biggl(  A \log({{M_{\rm string}^2}
\over {\mu ^2}}) +\sum_{\alpha=1}^3  {A'}^\alpha
\log\lbrack(t_\alpha+\bar t_\alpha)|\Delta (t_\alpha)|^2\rbrack\biggr),\cr
{1\over {\alpha _3(\mu )}}& = {3\over 8}
\biggl({1\over {\alpha_e(\mu)}} -{1\over {4\pi }} B
\log({{M_{\rm string}^2}\over {\mu ^2}})+
{{1}\over {4\pi }}\sum_{\alpha=1}^3{B'}^\alpha
\log\lbrack(t_\alpha+\bar t_\alpha)
|\Delta(t_\alpha)|^2\rbrack\biggr),\cr}  \eqn \alfs
$$
where $A
={k_2\over k_1}b_1-b_2$ and
$B= b_1+b_2-{k_1+k_2\over k_3}b_3$.
${A'}^\alpha$ and ${B'}^\alpha$
have the same expressions after replacing
$b_a\rightarrow {b'}_a^\alpha$.
For the MSSC
one has $A=28/5$ and
$B=20$.
Now we assume without loss of generality that the
threshold contributions dominantly come from one modulus, i.e.
from $t_1$, which belongs to a unrotated plane. Then one can
eliminate the explicit $t_1$ dependence from eqs.\alfs, and
with the experimental values for $\alpha_3^{\rm exp}$ and
$\sin^2\theta_W^{\rm exp}$
we obtain the following
condition on the modular weights of the standard model particles:
$2.7\leq
{{{B'}^1}/ {{A'}^1}} \leq3.7$.
Additional information about the allowed values of ${A'}^1$ and
${B'}^1$ can be extracted from the explicit $t_1$ dependence
of $\Delta(t_1)$. Specifically, if $\Delta(t_1)=\eta(t_1)^2$ one
knows that
$\log\lbrack(t_1+
\bar t_1)|\eta(t_1)|^4\rbrack<0$ for all possible values of $t_1$.
For a general threshold correction $\Delta(t_1)$ this inequality
is expected to hold,
at least for sufficiently large values of ${\rm Re}t_1$, which is
anyway required for these corrections to be large. This
originates from the expected  behavior of $\Delta(t_1)\rightarrow
e^{-t_1}$ for large ${\rm Re}t_1$, i.e. the Kaluza-Klein limit
of orbifold compactifications.
Then
the correct values low energy parameters are obtained provided
${A'}^1<0$, ${B'}^1<0$.
Checking these conditions on the modular weights
and taking into account also the
duality anomaly conditions for completely rotated planes
we obtain the following result:
The correct unification of the three gauge coupling constants
is possible for
the ${\bf Z}_6$, ${\bf Z}_8'$ and all ${\bf Z}_M\times{\bf Z}_N$
orbifolds. Large enough threshold correction require for these
cases that the radius of the relevant complex plane is
relatively large: ${\rm Re}t_1\sim 10 - 20$.

\subsec{Soft SUSY-breaking Parameters}

We will now turn to the phenomenology of the soft SUSY-breaking
parameters in four-dimensional strings
(see also ref.\JL ).
The presence of these soft terms reflects itself into
the SUSY-particle mass spectrum at low energies. If the idea of
low-energy supersymmetry is correct, the latter should be
amenable to experimental tests in future accelerators.
In principle there are as many different soft terms as independent
particles and/or couplings     present. Imposing some symmetries
at the GUT/Planck scale
reduces the number of independent soft terms.
In particular, it would
thus be very important  to find constraints on the pattern of
SUSY-breaking soft terms in effective low-energy
lagrangians from strings.
We will show that even without knowing the details of the
supersymmetry breaking process one can obtain
some characteristic features of the soft terms in a model independent
way. We will only assume that
the ``seed'' for SUSY-breaking
is provided by the auxiliary fields of the dilaton $s$ and the
moduli $t_\alpha$;
this assumption is true in most supersymmetry breaking
scenarios discussed up to now.

Let us consider first the gaugino masses for canonically
normalized gauginos
at the weak scale: $M_a(M_W)=2\pi\alpha_a(M_W)\tilde M_a$.
The value of $\tilde M_a$ is given in a general $N=1$ supergravity
lagrangian by\CREMMER
$$
\tilde M_a\ =\ m_{3/2}\sum _{\alpha=s,t}\ f_a^{\alpha }
\ K_{\alpha\bar\beta}^{-1}
\ G_{\bar\beta} . \eqn \mmm  $$
$m_{3/2}$ is the gravitino mass,
$f_a^\alpha$ is the derivative of the gauge kinetic function $f_a$
with respect to the  fields $s$ and $t _\alpha$,
$K_{\alpha\beta}^{-1}$
is the inverse K\"ahler metric and $G_\alpha$
is the auxiliary field
of $s$ and $t _\alpha$.
The gauge kinetic function has the general form
$$f_a(s,t_\alpha)=k_as+{c\over 16\pi^2}\log
\Delta(t_\alpha)^2.\eqn\gaugekina
$$
Here, $\log\Delta(t_\alpha)^2$ is the one-loop threshold contribution
from the massive string excitations given in terms of
automorphic functions of the corresponding duality  group $\Gamma$.
The constant $c$ is generically of order one.
As it stands, eq.\mmm, using eq.\gaugekina, only takes
into account the one-loop contribution of the massive fields,
however not the one-loop contribution of the massless fields in the
Yang-Mills lagrangian which is described by the non-local
effective langrangian \nl. Considering also this non-local
interaction simply amounts to replace
$f_a^{t_\alpha}$ by the derivative of the one-loop gauge
coupling constant with respect to $t_\alpha$\IL. This means that one
effectively has to replace in eq.\gaugekina\ $\Delta(t_\alpha)$
by
a non-holomorphic function $\tilde\Delta(t_\alpha,\bar t_\alpha)$
which contains also the anomalous piece of the massless fields.
Note that this contribution is also required by the duality
invariance of the gaugino mass.
Then $\tilde M_a$ takes the form
$$\tilde M_a=m_{3/2}\biggl(k_a
(K_{s\bar s}^{-1}G_{\bar s}+K_{s\bar t_\alpha}^{-1}G_{\bar t_\alpha}
)
+{1\over 8\pi^2\tilde\Delta}
{\partial\tilde\Delta_a\over\partial t_\alpha}
(K_{t_\alpha\bar s}^{-1}G_{\bar s}
+K_{t_\alpha\bar t_\beta}^{-1}G_{\bar t_\beta})\biggr)
{}.
\eqn
\gaugmassa
$$
Here we have allowed for a mixing between the $s$-field and the
moduli in the kinetic energy which occurs beyond the
tree-level\DFKZ.
Eq.\gaugmassa\ shows some interesting model-independent
features.
The existence of the threshold corrections implies in general
that the gaugino masses $\tilde M_a$
are non-universal\IL, i.e. gauge group dependent.
In particular, if supersymmetry breaking occurs mainly in the
moduli sector, i.e.
$G_{t_\alpha}>>G_s$,
which is often true for supersymmetry
breaking by gaugino condensation\GAUGINO\
in the hidden gauge sector,
the gaugino masses dominantly originate
from string loop effects. This means that $M_a$ is small
compared to $m_{3/2}$\FILQ\cc\CASAS.

For orbifolds one can explicitly parametrize the non-universality
of the gaugino masses by the
duality anomaly coefficients ${b'}_a^\alpha$.
With eq.\running\ one obtains
$$
\tilde M_a = M^0(s,t)k_a + \sum _{\alpha=1}^3
{b'}_a^\alpha {M'}^\alpha(s,t).
\eqn \mmb
$$
Here ${M'}^\alpha$
is due to loop effects and is small compared to $m_{3/2}$.
The size of the gauge group independent part $M^0$ depends
on the details of the supersymmetry breaking mechanism.
Generically it is of order $m_{3/2}$, however for
$G_{t_\alpha}>>G_s$ it is of one loop order and therefore
comparable to ${M'}^\alpha$.
It is instructive to eliminate the first term in the rhs. of \mmb\
by taking certain linear combinations of gaugino masses. This
leads to the following sum rule:
$$
\tilde M_1 (1-\gamma _M{{k_2}\over {k_1}}) + \tilde M_2 (1+\gamma _M)
 - {{k_1+k_2}\over {k_3}} \tilde M_3 = 0 .
\eqn \msem
$$
Here $
{\gamma }_M=(
{{\sum _{\alpha=1}^3\ {B'}^\alpha {M'}^\alpha})
/({\sum _{\alpha=1}^3\ {A'}^\alpha
{M'}^\alpha}})$ and
${A'}^\alpha$ and ${B'}^\alpha$ are the linear combinations introduced
in the previous section. Further simplification emerges (i)
if for the particular orbifold considered only the first plane
is left
unrotated by some twist. Then only ${M'}^1$ will be non-vanishing.
(ii) Or if the supersymmetry-breaking dynamics are such that
the modulus of a particular complex plane plays a leading role
(i.e. $|G_{t_1}|\gg |G_{t_2}|,|G_{t_3}|$), again only
${M'}^1$  will be relevant.
In both cases the moduli dependence cancels from $\gamma_M$ yielding
$\gamma_M\equiv\gamma={B'}^1/{A'}^1$.
In the MSSC scheme
we need to have
$2.7\leq \gamma \leq 3.7$ for the correct unification of the gauge
coupling constants. Then eq.\msem\ provides a rather
strong constraint on the gaugino masses.

Finally let us briefly display the general form of
the soft SUSY-breaking scalar masses.
The scalar potential in the effective low-energy supergravity
action has the form\CREMMER
$$ V =  e^G \biggl\lbrace
\sum_{i,j}         G_{\phi_i}
G_{\bar\phi_j}  K_{\phi_i\bar\phi_j}^{-1}
        - 3
  \biggr\rbrace ,\eqn\scalpotmatta
$$
where $K$ is the total K\"ahler potential and sum runs over
all charged massless chiral fields $\phi_i$.
With the K\"ahler potential \kporbi\ one gets
a general expression for the soft scalar masses of the form
$$
m_i^2=m_0^2(s,t_\alpha,\bar t_\alpha)+\sum_{\alpha=1}^3n^\alpha_i
{m'}^2(s,t_\alpha,\bar t_\alpha) .
\eqn \semesc
$$
The first term in the right-hand side is universal, i.e. does not
depend on the particular matter field $\phi_i$ considered.
The second term in eq.\semesc\ depends on the modular weights
of the matter fields
and is in general not
universal.
Interesting phenomenological constraints on the modular weights
may arise from
absence of flavor-changing neutral currents
demanding for example
that $m_{\tilde u}^2$ and $m_{\tilde c}^2$  must be almost degenerate
for the SUSY-GIM mechanism to work. This would suggest that
both fields have similar
modular weights. However for the case of gaugino condensation,
the considerations in ref.\CASAS\ indicate, that
${m'}^2\ll m_0^2$
such that the constraints on the modular weights might not be very tight.

\newsec{Conclusions}

In this talk we have considered some string consistency
as well as phenomenological constraints on four-dimensional strings.
One of the merits of
our argumentation is that it allows us to discard large classes
of models without needing to work on a (hopeless)
model-by-model basis.
Specifically we have shown that the requirement of the absence
of duality anomalies
rules out the existence of ${\bf Z}_3$ and ${\bf Z}_7$ orbifolds
(at the lowest Kac-Moody level) with the massless spectrum of
the supersymmetric standard model.
We believe that absence of duality anomalies
is intimately related to the world-sheet modular
invariance of four-dimensional strings, in complete analogy
to the relation\SCHEWA\ between gauge/gravitational anomalies and
world-sheet modular invariance.

Demanding for the additional, phenomenological
requirement of correct gauge coupling unification
all  ${\bf Z}_M$
orbifolds except ${\bf Z}_6$ and ${\bf Z}_8'$
are ruled out
when one assumes
the particle content of the minimal supersymmetric standard model.
On the other hand, a consistent MSSC
is not excluded (but not yet found)
for ${\bf Z}_M\times {\bf Z}_N$.
Finally we have shown that
in generic string models the soft scalar masses are $not$ universal
but depend on the modular weight, or in more general terms, on the
kinetic energy of the matter fields. Similarly,
one sees that the soft gaugino masses
are gauge group dependent. The departure from universality
of gaugino masses may be related in specific models to
the gauge coupling constant threshold effects. In some
cases specific mass relationships are found.

\bigskip
We like to thank L. Ib\'a\~nez for a most enjoyable collaboration
on the material presented in this review. Furthermore
we are very grateful to the organizers of the workshop for
making this stimulating meeting possible.

\vskip 0.3truein
\refout

\vfill
\eject
\end